\documentclass[technote]{IEEEtran}
\usepackage{amssymb}
\usepackage{algorithm}
\usepackage{algorithmic}
\usepackage{amsfonts}
\usepackage{amsmath}
\usepackage{amssymb}
\usepackage{amsthm}
\usepackage{array}
\usepackage{bbding}
\usepackage{bigints}
\usepackage{booktabs}
\usepackage{cite}
\usepackage{cleveref}
\usepackage{color}
\usepackage{diagbox}
\usepackage{epsfig,latexsym}
\usepackage{epstopdf}
\usepackage{graphicx}
\usepackage{fancyhdr}
\usepackage{float}
\usepackage{flushend}
\usepackage{indentfirst}
\usepackage{lastpage}
\usepackage{makecell}
\usepackage{mathtools}
\usepackage{multirow}
\usepackage{pifont}
\usepackage{psfrag}
\usepackage{setspace}
\usepackage{stfloats}
\usepackage{subfloat}
\usepackage{times}
\usepackage{subfig}

\theoremstyle{remark}
\newtheorem{theorem}{\quad \textbf{Theorem}}
\newtheorem{lemma}{\quad \textbf{Lemma}}

\newtheorem{remark}{\quad \textbf{Remark}}

\allowdisplaybreaks[4]

\begin{document}
\title{Achievable Rate Analysis of Intelligent Omni-Surface Assisted NOMA Holographic MIMO Systems}

\author{Qingchao Li, \textit{Graduate Student Member, IEEE}, \\Mohammed El-Hajjar, \textit{Senior Member, IEEE},\\ Yanshi Sun, \textit{Member, IEEE}, \\Ibrahim Hemadeh, \textit{Senior Member, IEEE},\\ Yingming Tsai, \textit{Member, IEEE}, \\Arman Shojaeifard, \textit{Senior Member, IEEE},\\ and Lajos Hanzo, \textit{Life Fellow, IEEE}

\thanks{\textit{(Corresponding author: Lajos Hanzo.)}

Qingchao Li, Mohammed El-Hajjar and Lajos Hanzo are with the School of Electronics and Computer Science, University of Southampton, Southampton SO17 1BJ, U.K. (e-mail: Qingchao.Li@soton.ac.uk; meh@ecs.soton.ac.uk; lh@ecs.soton.ac.uk).

Yanshi Sun is with the School of Computer Science and Information Engineering, Hefei University of Technology, Hefei, 230009, China. (email:sys@hfut.edu.cn).

Ibrahim Hemadeh, Yingming Tsai and Arman Shojaeifard are with InterDigital, London EC2A 3QR, U.K. (e-mail: Ibrahim.Hemadeh@interdigital.com; Yingming.Tsai@interdigital.com; Arman.Shojaeifard@interdigital.com).}}

\maketitle

\begin{abstract}
An intelligent omni-surface (IOS) assisted holographic multiple-input and multiple-output architecture is conceived for $360^\circ$ full-space coverage at a low energy consumption. The theoretical ergodic rate lower bound of our non-orthogonal multiple access (NOMA) scheme is derived based on the moment matching approximation method, while considering
the signal distortion at transceivers imposed by hardware impairments (HWIs). Furthermore, the asymptotically ergodic rate lower bound is derived both for an infinite number of IOS elements and for continuous aperture surfaces. Both the theoretical analysis and the simulation results show that the achievable rate of the NOMA scheme is higher than that of its orthogonal multiple access counterpart. Furthermore, owing to the HWIs at the transceivers, the achievable rate saturates at high signal-to-noise ratio region, instead of reaching its theoretical maximum.
\end{abstract}

\section{Introduction}
Holographic multiple-input and multiple-output (HMIMO) systems are expected to evolve towards an intelligent software reconfigurable paradigm in support of improved spectral efficiency. They can be realized by harnessing a large number antennas for constructing a spatially near-continuous aperture~\cite{huang2020holographic}. However, it is infeasible to realize HMIMO schemes relying a large number of conventional radio frequency (RF) chains and antennas due to the excessive power consumption.

In~\cite{deng2021reconfigurable_wc,deng2021reconfigurable_tvt}, Deng \textit{et al.} proposed a reconfigurable holographic surface (RHS), consisting of multiple feeds and a large number of metamaterial based radiation elements. They constructed a hybrid beamforming scheme, where the digital beamformer and the holographic beamformer are employed at the base station (BS) and the RHS, respectively. To be specific, the digital beamformer relies on the state-of-the-art zero-forcing transmit precoding method, while the holographic beamformer relies on the coefficient configuration of the amplitude-controlled RHS radiation elements. The digital beamformer and the holographic beamformer are alternatively optimized for maximizing the achievable sum-rate. The simulation results showed that the RHS assisted hybrid beamformer achieves higher sum-rate than the state-of-the-art massive MIMO hybrid beamformer based on phase shift arrays. Furthermore, Hu \textit{et al.}~\cite{hu2022holographic} proposed a holographic beamformer based on amplitude-controlled RHS elements having a finite resolution. They demonstrated that the holographic beamformer associated with as few as 2-bit quantization approached the sum-rate associated with unquantized values.

To further increase the energy efficiency, Zeng \textit{et al.} in~\cite{zeng2022reconfigurable} conceived a single-RF based HMIMO architecture, where a reconfigurable refractive surface (RRS) illuminated by a single feed is employed at the BS for beamforming by optimizing the coefficient of each RRS element. Both the theoretical analysis and the simulation results demonstrate that at the same data rate as the conventional MIMO systems relying on phased arrays, the RRS-based HMIMO systems have higher energy efficiency. Nevertheless, in the RRS-assisted HMIMO architecture, only the users in the $180^\circ$ half-plane can be supported. In~\cite{wu2021coverage}, \cite{yue2022simultaneously}, \cite{wu2022resource}, the technique of intelligent omni-surface (IOS), also termed as simultaneously transmitting and reflecting reconfigurable intelligent surface, was introduced to realize $360^\circ$ full-space coverage. Furthermore, it was demonstrated that the outage probability performance of the IOS assisted non-orthogonal multiple access (NOMA) systems outperforms that of the IOS assisted orthogonal multiple access (OMA) systems. However, idealized perfect hardware quality was assumed at the RF-chains of transceivers in~\cite{deng2021reconfigurable_wc}, \cite{deng2021reconfigurable_tvt}, \cite{hu2022holographic},
\cite{zeng2022reconfigurable}, \cite{wu2021coverage}, \cite{yue2022simultaneously}, \cite{wu2022resource}, ignoring the signal distortion resulting from practical hardware impairments (HWIs). To deal with these issues, in this paper we propose an IOS assisted HMIMO architecture, while considering the effect of HWIs, where the modulated signal can be both reflected and refracted from the IOS, to realize $360^\circ$ full-space coverage. Our contributions in this compact letter are presented as follows:

\begin{itemize}
  \item We conceive an IOS assisted HMIMO architecture having low energy consumption and $360^\circ$ full-space coverage. The theoretical lower bound of the achievable ergodic rate of our NOMA scheme relying on the popular successive interference cancellation (SIC) based reception is derived based on the moment matching approximation method. Additionally, the distortion of both the signal emission at the transmitters and the signal reception at the receivers imposed by the HWIs is also taken into account. Furthermore, the asymptotic ergodic rate lower bound is derived both when the number of IOS elements tends to infinity and for continuous aperture surfaces.
  \item The theoretical analysis and simulation results show that the achievable rate of our NOMA scheme is higher than that of its OMA counterpart in the IOS assisted HMIMO systems. Furthermore, owing to the HWIs of transceivers, the achievable rate saturates at high signal-noise ratios (SNR) instead of reaching its theoretical maximum.
\end{itemize}

\textit{Notations:} $\jmath=\sqrt{-1}$, $|a|$ represents the amplitude of the complex scalar $a$, $\mathcal{CN}\left(\mu,\sigma^2\right)$ is a circularly symmetric complex Gaussian random variable with the mean $\mu$ and the variance $\sigma^2$, $\mathbb{E}\left[x\right]$ represents the mean of the random variable $x$, $\mathop{\sum_{(n_1,n_2,\cdots,n_q)=1}^{N}}$ represents $\mathop{\sum_{n_1=1}^{N}\sum_{n_2=1}^{N}\cdots\sum_{n_q=1}^{N}}\limits_{n_1\neq n_2\neq\cdots\neq n_q}$.

\section{System Model}\label{System_Model}
The downlink of our holographic MIMO system is illustrated in Fig.~\ref{Fig_system_model_IOS}, where an intelligent omni-surface illuminated by a single RF-chain feed is used at the BS for beamforming in support of two single-antenna user equipment (UE) at different sides of the IOS. In contrast to~\cite{zeng2022reconfigurable}, where a reconfigurable refractive surface is employed at the BS and only the users at one side of the surface can be supported, the intelligent omni-surface considered achieves $360^\circ$ full-space coverage. We denote the user located within the reflecting area as UE-1 and the user in the refracting area as UE-2.

\subsection{Intelligent Omni-Surface}
\begin{figure}[!t]
    \centering
    \includegraphics[width=3in]{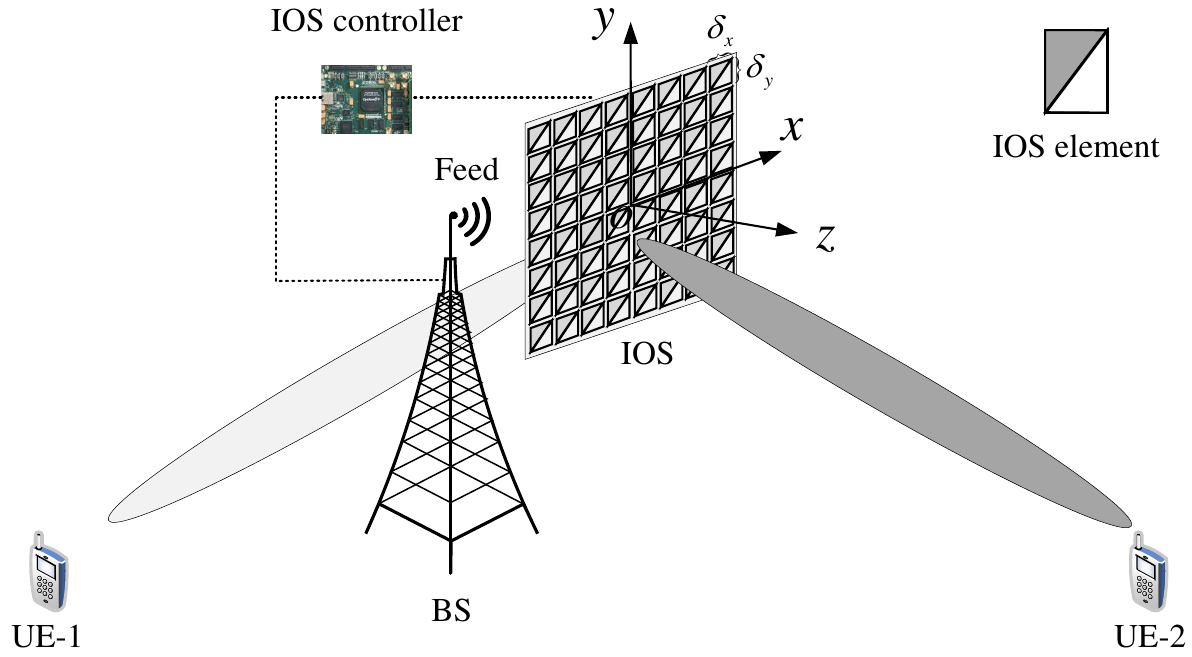}
    \caption{System model of the IOS assisted HMIMO architecture.}\label{Fig_system_model_IOS}
\end{figure}

As shown in Fig.~\ref{Fig_system_model_IOS}, the $xoy$ plane coincides with the IOS and the origin is located at the center of the IOS. We assume that a total of $N=N_x\times N_y$ IOS elements are compactly placed in a uniform rectangular planar array, with $N_x$ and $N_y$ elements in the $x$ axis direction and $y$ axis direction, respectively. Each IOS element has the size of $\delta_x$ and $\delta_y$ in the $x$ axis direction and $y$ axis direction, respectively. To arrange for the intelligent surface to both reflect and refract the impinging signal simultaneously, the energy splitting protocol of~\cite{zhang2022star} is employed. Specifically, each IOS element can split the total energy between the reflection and refraction modes. The coefficient of the $n$th IOS element in the reflection mode and refraction mode can be represented as $\zeta_{1,n}=\beta_{1,n}\mathrm{e}^{\jmath\theta_{1,n}}$ and $\zeta_{2,n}=\beta_{2,n}\mathrm{e}^{\jmath\theta_{2,n}}$, respectively, where it must be satisfied that $\beta_{1,n}^2+\beta_{2,n}^2=1$. Referring to~\cite{zhao2022ergodic}, we can set $\beta_{1,1}=\beta_{1,2}=\cdots=\beta_{1,N}=\beta_1$ and $\beta_{2,1}=\beta_{2,2}=\cdots=\beta_{2,N}=\beta_2$ for reasons of hardware simplicity. Thus, for the $n$th IOS element we have:
\begin{align}\label{System_Model_1}
    \zeta_{i,n}=\beta_i\mathrm{e}^{\jmath\theta_{i,n}},\quad i\in\{1,2\},
\end{align}
subject to $\beta_1^2+\beta_2^2=1$.

\subsection{Channel Model}
\subsubsection{Channel link spanning from the feed to the IOS}
Since the feed is near the IOS, the near-field channel model is employed for the link spanning from the feed to the IOS. We assume that the feed is located on the normal of the IOS through the origin with the distance of $d_0$. Hence the coordinates of the feed is $\left(0,0,-d_0\right)$. Besides, we denote the coordination of the center point of the $n$th IOS element as $\left(x_n,y_n,0\right)$. Therefore, the distance between the feed and the center point of the $n$th IOS element can be expressed as
\begin{align}\label{Channel_Model_1}
    d_n=\sqrt{x_n^2+y_n^2+d_0^2},
\end{align}
while the energy impinging on the $n$th IOS element can be represented as \cite{zeng2022reconfigurable}
\begin{align}\label{Channel_Model_2}
    \gamma_n=\iint_{\mathcal{D}_n}\frac{\left(\alpha+1\right)d_0^{\alpha+1}}{2\pi}
    \left(d_0^2+x^2+y^2\right)^{-\frac{\alpha+3}{2}}\mathrm{d}x\mathrm{d}y,
\end{align}
where $2\left(\alpha+1\right)$ is the gain of the feed, and the integration region is $\mathcal{D}_n=\{\left(x,y\right)\in\mathbb{R}^2:x_n-\frac{\delta_x}{2}\leq x\leq x_n+\frac{\delta_x}{2},y_n-\frac{\delta_y}{2}\leq y\leq y_n+\frac{\delta_y}{2}\}$.
According to (\ref{Channel_Model_1}) and (\ref{Channel_Model_2}), we can describe the channel link spanning from the feed to the $n$th IOS element, denoted as $g_n$, as
\begin{align}\label{Channel_Model_3}
    g_n=\sqrt{\gamma_n}\mathrm{e}^{-\jmath\frac{2\pi}{\lambda}d_n}
    =\sqrt{\gamma_n}\mathrm{e}^{-\jmath\frac{2\pi}{\lambda}\sqrt{x_n^2+y_n^2+d_0^2}},
\end{align}
where $\lambda$ is the carrier wavelength.

\subsubsection{Channel link spanning from the IOS to UEs}
We consider the far-field Nakagami-$m$ fading channel model for the channel link spanning from the IOS to the UEs. The channel links spanning from the $n$th IOS element to UE-1 and UE-2 are given by
\begin{align}\label{Channel_Model_4}
    h_{i,n}=\sqrt{\varrho_i}q_{i,n}\mathrm{e}^{\jmath\psi_{i,n}},\quad i\in\{1,2\},
\end{align}
where $\varrho_i$ represents the large-scale fading between the IOS to UE-$i$, and $q_{i,n}\mathrm{e}^{\jmath\psi_{i,n}}$ is the small-scale fading. Specifically, in the small-scale fading, the phase shift $\psi_{i,n}$ obeys the identical uniform distribution within the support interval of $[0,2\pi)$. The amplitude $q_{i,n}$ follows the Nakagami distribution having the probability density function of $f_i(x)=\frac{2m_i^{m_i}}{\Gamma\left(m_i\right)}x^{2m_i-1}\mathrm{e}^{-m_ix^2}$, where $\Gamma\left(\cdot\right)$ represents the gamma function while $m_i$ is the shape parameter in the corresponding link.

\subsubsection{Equivalent channel from the feed to UEs}
According to (\ref{System_Model_1}), (\ref{Channel_Model_3}) and (\ref{Channel_Model_4}), we can characterize the equivalent channel spanning from the feed to UE-$i$ as $h_i=\sum_{n=1}^{N}g_n\zeta_{i,n}h_{i,n}=\sqrt{\varrho_i}\beta_i\sum_{n=1}^{N}
\sqrt{\gamma_n}q_{i,n}\mathrm{e}^{-\jmath\frac{2\pi}{\lambda}\sqrt{x_n^2+y_n^2+d_0^2}}
\mathrm{e}^{\jmath\theta_{i,n}}\mathrm{e}^{\jmath\psi_{i,n}}$. To maximize the channel gain of the equivalent channels, the phase shift of the IOS element can be configured as $\theta_{i,n}=\frac{2\pi}{\lambda}\sqrt{x_n^2+y_n^2+d_0^2}-\psi_{i,n}$. Therefore, the equivalent channel between the feed and UE-1 as well as UE-2 are given by
\begin{align}\label{Channel_Model_8}
    h_i=\sqrt{\varrho_i}\beta_i\sum_{n=1}^{N}\sqrt{\gamma_n}q_{i,n},\quad i\in\{1,2\}.
\end{align}

\subsection{NOMA Design}
Since the NOMA principle can serve additional far-field users by exploiting the spatial beams precoded for the legacy near-field users~\cite{yue2022simultaneously}, we focus our attention on the NOMA architecture, where the BS uses superposition coding for non-orthogonal multiplexing and the UEs employ the classic SIC algorithm for recovering the information \cite{dai2018survey}. We denote the signal for UE-1 and UE-2 as $s_1$ and $s_2$ respectively, and the transmit signal at the feed can be formulated as $s=\sqrt{\kappa_1}s_1+\sqrt{\kappa_2}s_2$, where $\kappa_1$ and $\kappa_2$ are the power allocation coefficients satisfying $\kappa_1+\kappa_2=1$. Thus, the signals received by UE-1 and UE-2 are \cite{li2023performance_tcom}
\begin{align}\label{NOMA_Design_1}
    \notag y_1=&\underbrace{\sqrt{\rho \kappa_1\varepsilon_{u^{(1)}}\varepsilon_v}h_1s_1}
    _{\text{Desired signal of $s_1$}}
    +\underbrace{\sqrt{\rho \kappa_1\varepsilon_{u^{(1)}}\left(1-\varepsilon_v\right)}h_1v_1}
    _{\text{BS HWI distortion on $s_1$}}\\
    \notag&+\underbrace{\sqrt{\rho\kappa_1\left(1-\varepsilon_{u^{(1)}}\right)}h_1u^{(1)}_1}
    _{\text{UE-1 HWI distortion on $s_1$}}+\underbrace{\sqrt{\rho \kappa_2\varepsilon_{u^{(1)}}\varepsilon_v}h_1s_2}
    _{\text{Desired signal of $s_2$}}\\
    \notag&+\underbrace{\sqrt{\rho \kappa_2
    \varepsilon_{u^{(1)}}\left(1-\varepsilon_v\right)}h_1v_2}
    _{\text{BS HWI distortion on $s_2$}}+\underbrace{\sqrt{\rho \kappa_2\left(1-\varepsilon_{u^{(1)}}\right)}h_1u^{(1)}_2}
    _{\text{UE-1 HWI distortion on $s_2$}}\\
    &+\underbrace{w_1}_{\text{Additive noise at UE-1}},
\end{align}
and
\begin{align}\label{NOMA_Design_2}
    \notag y_2=&\underbrace{\sqrt{\rho \kappa_1\varepsilon_{u^{(2)}}\varepsilon_v}h_2s_1}
    _{\text{Desired signal of $s_1$}}+\underbrace{\sqrt{\rho \kappa_1
    \varepsilon_{u^{(2)}}\left(1-\varepsilon_v\right)}h_2v_1}_{\text{BS HWI distortion on $s_1$}}\\
    \notag&+\underbrace{\sqrt{\rho \kappa_1\left(1-\varepsilon_{u^{(2)}}\right)}h_2u^{(2)}_1}
    _{\text{UE-2 HWI distortion on $s_1$}}+\underbrace{\sqrt{\rho \kappa_2\varepsilon_{u^{(2)}}\varepsilon_v}h_2s_2}
    _{\text{Desired signal of $s_2$}}\\
    \notag&+\underbrace{\sqrt{\rho \kappa_2
    \varepsilon_{u^{(2)}}\left(1-\varepsilon_v\right)}h_2v_2}
    _{\text{BS HWI distortion on $s_2$}}+\underbrace{\sqrt{\rho \kappa_2\left(1-\varepsilon_{u^{(2)}}\right)}h_2u^{(2)}_2}
    _{\text{UE-2 HWI distortion on $s_2$}}\\
    &+\underbrace{w_2}_{\text{Additive noise at UE-2}},
\end{align}
where $\rho$ denotes the total transmit power at the BS. Furthermore, $v_1\sim\mathcal{CN}\left(0,1\right)$, $u^{(1)}_1\sim\mathcal{CN}\left(0,1\right)$ and $u^{(2)}_1\sim\mathcal{CN}\left(0,1\right)$ are the distortion of the information symbol $s_1$ due to hardware impairments, resulting from power amplifier non-linearities, amplitude/phase imbalance in the In-phase/Quadrature mixers, phase noise in the local oscillator, sampling jitter and finite-resolution quantization in the analog-to-digital converters \cite{bjornson2017massive}, at the BS, UE-1 and UE-2, respectively. Similarly, $v_2\sim\mathcal{CN}\left(0,1\right)$, $u^{(1)}_2\sim\mathcal{CN}\left(0,1\right)$ and $u^{(2)}_2\sim\mathcal{CN}\left(0,1\right)$ are the distortion of the information symbol $s_2$ due to hardware impairments at the BS, UE-1 and UE-2, respectively. Finally, $\varepsilon_v$, $\varepsilon_{u^{(1)}}$ and $\varepsilon_{u^{(2)}}$ represent the hardware quality factor of the BS, UE-1 and UE-2, respectively, satisfying $0\leq\varepsilon_v\leq1$, $0\leq\varepsilon_{u^{(1)}}\leq1$ and $0\leq\varepsilon_{u^{(2)}}\leq1$. Explicitly, a hardware quality factor of 1 indicates that the hardware is perfectly ideal, while 0 means that the hardware is completely inadequate \cite{li2023achievable_tvt}. Furthermore, $w_1\sim\mathcal{CN}\left(0,\sigma_{w_1}^2\right)$ and $w_2\sim\mathcal{CN}\left(0,\sigma_{w_2}^2\right)$ represent the additive noise at UE-1 and UE-2, respectively.

At the user side, we employ the SIC algorithm for recovering information. Without loss of generality, we assume that the channel gain of UE-1 is higher than that of UE-2, i.e. $\frac{\left|h_1\right|^2}{\sigma_{w_1}^2}>\frac{\left|h_2\right|^2}{\sigma_{w_2}^2}$. In this case, UE-2 extracts its signal $s_2$ from the received composite signal $y_2$ by regarding the signal of UE-1 as interference. By contrast, UE-1 firstly detects the signal of $s_2$, and then subtracts its remodulated version from the received composite signal $y_1$, so that UE-1 can detect its own signal $s_1$ without interference from UE-2. Therefore, according to (\ref{NOMA_Design_1}) and (\ref{NOMA_Design_2}), the achievable rate of UE-1 and UE-2 can be formulated as $R_1=\log_2\big(1+\frac{\rho\kappa_1\varepsilon_{u^{(1)}}\varepsilon_v\left|h_1\right|^2}
{\rho\kappa_1(1-\varepsilon_{u^{(1)}}\varepsilon_v)\left|h_1\right|^2+
\rho\kappa_2(1-\varepsilon_{u^{(2)}}\varepsilon_v)\left|h_1\right|^2+\sigma_{w_1}^2}\big)$ and $R_2=\log_2\left(1+\frac{\rho \kappa_2\varepsilon_{u^{(2)}}\varepsilon_v\left|h_2\right|^2}
{\rho \kappa_2\left(1-\varepsilon_{u^{(2)}}\varepsilon_v\right)\left|h_2\right|^2+\rho \kappa_1\left|h_2\right|^2+\sigma_{w_2}^2}\right)$, respectively.

For comparison, the achievable rate of UE-1 and UE-2 in the OMA protocol can be expressed as $R_1'=\kappa_1'\log_2\Big(1+\frac{\rho \varepsilon_{u^{(1)}}\varepsilon_v\left|h_1\right|^2}
{\rho\left(1-\varepsilon_{u^{(1)}}\varepsilon_v\right)\left|h_1\right|^2+\sigma_{w_1}^2}\Big)$ and $R_2'=\kappa_2'\log_2\Big(1+\frac{\rho \varepsilon_{u^{(2)}}\varepsilon_v\left|h_2\right|^2}
{\rho\left(1-\varepsilon_{u^{(2)}}\varepsilon_v\right)\left|h_2\right|^2+\sigma_{w_2}^2}\Big)$, respectively, where $\kappa_1'$ and $\kappa_2'$ are the orthogonal time/frequency resource ratio allocated to UE-1 and UE-2 respectively, satisfying $\kappa_1'+\kappa_2'=1$.

\section{Performance Analysis}\label{Performance_Analysis}
In this section, we theoretically derive the ergodic rate of the NOMA scheme.

The ergodic rate of UE-1 and UE-2 can be formulated as
\begin{align}\label{NOMA_Design_3}
    \notag&\mathbb{E}\left[R_1\right]=\mathbb{E}\Bigg[\log_2\Bigg(1+\\
    \notag&\frac{\rho\kappa_1\varepsilon_{u^{(1)}}\varepsilon_v\left|h_1\right|^2}
    {\rho\kappa_1\left(1-\varepsilon_{u^{(1)}}\varepsilon_v\right)\left|h_1\right|^2+
    \rho\kappa_2\left(1-\varepsilon_{u^{(2)}}\varepsilon_v\right)
    \left|h_1\right|^2+\sigma_{w_1}^2}\Bigg)\Bigg]\\
    =&\mathbb{E}\left[\log_2\left(1+\frac{\rho\kappa_1\varepsilon_{u^{(1)}}\varepsilon_v}
    {\rho\left(1-\kappa_1\varepsilon_{u^{(1)}}\varepsilon_v
    -\kappa_2\varepsilon_{u^{(2)}}\varepsilon_v\right)
    +\sigma_{w_1}^2\frac{1}{\left|h_1\right|^2}}\right)\right],
\end{align}
and
\begin{align}\label{NOMA_Design_4}
    \notag&\mathbb{E}\left[R_2\right]\\
    \notag=&\mathbb{E}\Bigg[\log_2\Bigg(1+
    \frac{\rho\kappa_2\varepsilon_{u^{(2)}}\varepsilon_v\left|h_2\right|^2}
    {\rho\kappa_2\left(1-\varepsilon_{u^{(2)}}\varepsilon_v\right)\left|h_2\right|^2+
    \rho\kappa_1\left|h_2\right|^2+\sigma_{w_2}^2}\Bigg)\Bigg]\\
    =&\mathbb{E}\left[\log_2\left(1+\frac{\rho\kappa_2\varepsilon_{u^{(2)}}\varepsilon_v}
    {\rho\left(1-\kappa_2\varepsilon_{u^{(2)}}\varepsilon_v\right)
    +\sigma_{w_2}^2\frac{1}{\left|h_2\right|^2}}\right)\right],
\end{align}
respectively. Based on (\ref{NOMA_Design_3}) and (\ref{NOMA_Design_4}), as well as the Jensen's inequality, the lower bound of the achievable ergodic rate of UE-1 and UE-2, denoted as $\overline{R}_1$ and $\overline{R}_2$ respectively, become
\begin{align}\label{NOMA_Design_5}
    \notag\overline{R}_1=&\log_2\Bigg(1+\\
    &\frac{\rho\kappa_1\varepsilon_{u^{(1)}}\varepsilon_v}
    {\rho\left(1-\kappa_1\varepsilon_{u^{(1)}}\varepsilon_v-\kappa_2\varepsilon_{u^{(2)}}
    \varepsilon_v\right)
    +\sigma_{w_1}^2\mathbb{E}\left[\frac{1}{\left|h_1\right|^2}\right]}\Bigg),
\end{align}
and
\begin{align}\label{NOMA_Design_6}
    \overline{R}_2=\log_2\left(1+\frac{\rho\kappa_2\varepsilon_{u^{(2)}}\varepsilon_v}
    {\rho\left(1-\kappa_2\varepsilon_{u^{(2)}}\varepsilon_v\right)
    +\sigma_{w_2}^2\mathbb{E}\left[\frac{1}{\left|h_2\right|^2}\right]}\right),
\end{align}
respectively. Therefore, to theoretically derive $\overline{R}_1$ and $\overline{R}_2$, our objective is to determine the values of $\mathbb{E}\left[\frac{1}{\left|h_1\right|^2}\right]$ and $\mathbb{E}\left[\frac{1}{\left|h_2\right|^2}\right]$, which can be derived based on the moment matching approximation method as follows.

\begin{lemma}\label{lemma_1}
According to (\ref{Channel_Model_8}), the first and second moment of $\left|h_i\right|^2$, denoted as $\mathbb{E}[\left|h_i\right|^2]$ and $\mathbb{E}[\left|h_i\right|^4]$ respectively, are
\begin{align}\label{NOMA_Design_7}
    \notag\mathbb{E}[\left|h_i\right|^2]
    =&\mathbb{E}\left[\left(\beta_i\sqrt{\varrho_i}\sum_{n=1}^{N}\sqrt{\gamma_n}q_{i,n}\right)^2\right]\\
    \notag=&\varrho_i\beta_i^2\left(\mu_{i,2}\sum_{n=1}^{N}\gamma_n
    +\mu_{i,1}^2\mathop{\sum_{(n_1,n_2)=1}^{N}}\sqrt{\gamma_{n_1}\gamma_{n_2}}\right)\\
    =&\varrho_i\beta_i^2\Big(\mu_{i,2}A_2+\mu_{i,1}^2\left(A_1^2-A_2\right)\Big),
\end{align}
and
\begin{align}\label{NOMA_Design_8}
    &\notag\mathbb{E}\left[\left|h_i\right|^4\right]
    =\mathbb{E}\left[\left(\beta_i\sqrt{\varrho_i}
    \sum_{n=1}^{N}\sqrt{\gamma_n}q_{i,n}\right)^4\right]\\
    \notag=&\varrho_i^2\beta_i^4\Bigg(\mu_{i,4}\sum_{n=1}^{N}\gamma_n^2
    +4\mu_{i,3}\mu_{i,1}\mathop{\sum_{(n_1,n_2)=1}^{N}}
    \gamma_{n_1}^{\frac{3}{2}}\gamma_{n_2}^{\frac{1}{2}}\\
    \notag&+3\mu_{i,2}^2\mathop{\sum_{(n_1,n_2)=1}^{N}}
    \gamma_{n_1}\gamma_{n_2}+6\mu_{i,2}\mu_{i,1}^2\mathop{\sum_{(n_1,n_2,n_3)=1}^{N}}
    \gamma_{n_1}\gamma_{n_2}^{\frac{1}{2}}\gamma_{n_3}^{\frac{1}{2}}\\
    \notag&+\mu_{i,1}^4\mathop{\sum_{(n_1,n_2,n_3,n_4)=1}^{N}}
    \gamma_{n_1}^{\frac{1}{2}}\gamma_{n_2}^{\frac{1}{2}}
    \gamma_{n_3}^{\frac{1}{2}}\gamma_{n_4}^{\frac{1}{2}}\Bigg)\\
    \notag=&\varrho_i^2\beta_i^4\Big(\mu_{i,4}A_4
    +4\mu_{i,3}\mu_{i,1}\left(A_3A_1-A_4\right)+3\mu_{i,2}^2\left(A_2^2-A_4\right)\\
    \notag&+6\mu_{i,2}\mu_{i,1}^2\left(A_2A_1^2-2A_3A_1-A_2^2+2A_4\right)+\\
    &\mu_{i,1}^4\left(A_1^4-6A_2A_1^2+3A_2^2+8A_3A_1-6A_4\right)\Big),
\end{align}
respectively, where the $k$th ($k=1,2,3,4$) moment of $q_{i,n}$ is given by $\mu_{i,1}=\mathbb{E}\left[q_{i,n}\right]=\frac{\Gamma\left(m_i+\frac{1}{2}\right)}{\Gamma(m_i)}
\left(\frac{1}{m_i}\right)^{\frac{1}{2}}$, $\mu_{i,2}=\mathbb{E}\left[q_{i,n}^2\right]=1$, $\mu_{i,3}=\mathbb{E}\left[q_{i,n}^3\right]=\frac{\Gamma\left(m_i+\frac{3}{2}\right)}{\Gamma(m_i)}
\left(\frac{1}{m_i}\right)^{\frac{3}{2}}$, and $\mu_{i,4}=\mathbb{E}\left[q_{i,n}^4\right]=1+\frac{1}{m_i}$. Furthermore, we have $A_1=\left(\frac{\delta_x\delta_y}{\lambda^2}\right)^{-\frac{1}{2}}
\iint_{\mathcal{D}}\omega^{\frac{1}{2}}\mathrm{d}x\mathrm{d}y$, $A_2=\iint_{\mathcal{D}}\omega\mathrm{d}x\mathrm{d}y$, $A_3=\left(\frac{\delta_x\delta_y}{\lambda^2}\right)^{\frac{1}{2}}
\iint_{\mathcal{D}}\omega^{\frac{3}{2}}\mathrm{d}x\mathrm{d}y$ and $A_4=\left(\frac{\delta_x\delta_y}{\lambda^2}\right)
\iint_{\mathcal{D}}\omega^{2}\mathrm{d}x\mathrm{d}y$, with $\omega=\frac{\left(\alpha+1\right)d_0^{\alpha+1}}{2\pi}\left(d_0^2+x^2+y^2\right)
^{-\frac{\alpha+3}{2}}$ and the integration region is given by $\mathcal{D}=\sum_{n=1}^{N}\mathcal{D}_n=\{\left(x,y\right)\in\mathbb{R}^2:
-\frac{N_x}{2}\delta_x\leq x\leq\frac{N_x}{2}\delta_x,
-\frac{N_y}{2}\delta_y\leq y\leq\frac{N_y}{2}\delta_y$\}.
\end{lemma}
\begin{IEEEproof}
    (\ref{NOMA_Design_7}) and (\ref{NOMA_Design_8}) can be derived based on the polynomial theorem and on the independence of the channel links spanning from each IOS element to the UEs.
\end{IEEEproof}

Based on the moment matching approximation method, the random variable of $\left|h_i\right|^2$ approximately follows the Gamma distribution having the shape parameter $\nu_i=\frac{(\mathbb{E}[\left|h_i\right|^2])^2}{\mathbb{E}[\left|h_i\right|^4]
-(\mathbb{E}[\left|h_i\right|^2])^2}$ and the scale parameter $\varsigma_i=\frac{\mathbb{E}[\left|h_i\right|^4]-(\mathbb{E}[\left|h_i\right|^2])^2}
{\mathbb{E}[\left|h_i\right|^2]}$, respectively. Therefore, $\frac{1}{\left|h_i\right|^2}$ follows the inverse gamma distribution associated with the shape parameter $\nu_i'=\nu_i=\frac{(\mathbb{E}[\left|h_i\right|^2])^2}{\mathbb{E}[\left|h_i\right|^4]
-(\mathbb{E}[\left|h_i\right|^2])^2}$ and the scale parameter $\varsigma_i'=\frac{1}{\varsigma_i}
=\frac{\mathbb{E}[\left|h_i\right|^2]}{\mathbb{E}[\left|h_i\right|^4]
-(\mathbb{E}[\left|h_i\right|^2])^2}$. Thus, the means of $\frac{1}{\left|h_i\right|^2}$ are given by
\begin{align}\label{NOMA_Design_13}
    \mathbb{E}\left[\frac{1}{\left|h_i\right|^2}\right]=\frac{\varsigma_i'}{\nu_i'-1}
    =\frac{\mathbb{E}[\left|h_i\right|^2]}{2(\mathbb{E}[\left|h_i\right|^2])^2
    -\mathbb{E}[\left|h_i\right|^4]},
\end{align}
with $\mathbb{E}\left[\left|h_i\right|^2\right]$ and $\mathbb{E}\left[\left|h_i\right|^4\right]$ given in (\ref{NOMA_Design_7}) and (\ref{NOMA_Design_8}), respectively. Furthermore, by substituting (\ref{NOMA_Design_13}) into (\ref{NOMA_Design_5}) and (\ref{NOMA_Design_6}), the theoretical values of $\overline{R}_1$ and $\overline{R}_2$ can be formulated as
\begin{align}\label{NOMA_Design_14_1}
    \overline{R}_1=\log_2\left(1+\frac{\rho\varrho_1\beta_1^2
    \kappa_1\varepsilon_{u^{(1)}}\varepsilon_v}{\rho\varrho_1\beta_1^2
    \left(1-\kappa_1\varepsilon_{u^{(1)}}\varepsilon_v-\kappa_2\varepsilon_{u^{(2)}}
    \varepsilon_v\right)+\eta_1\sigma_{w_1}^2}\right),
\end{align}
and
\begin{align}\label{NOMA_Design_14_2}
    \overline{R}_2=\log_2\left(1+\frac{\rho\varrho_2\beta_2^2\kappa_2
    \varepsilon_{u^{(2)}}\varepsilon_v}{\rho\varrho_2\beta_2^2
    \left(1-\kappa_2\varepsilon_{u^{(2)}}\varepsilon_v\right)+\eta_2\sigma_{w_2}^2}\right),
\end{align}
respectively, where $\eta_1$ and $\eta_2$ are given by
\begin{align}\label{NOMA_Design_15}
    \eta_i=\frac{\iota_{i,1}^2A_1^2+\iota_{i,2}A_2}{\iota_{i,1}^4A_1^4-2\iota_{i,2}
    \iota_{i,1}^2A_2A_1^2-\iota_{i,2}^2A_2^{2}-4\iota_{i,3}\iota_{i,1}A_3A_1-\iota_{i,4}A_4},
\end{align}
with $i\in\{1,2\}$ and
$\iota_{i,1}=\mu_{i,1}$, $\iota_{i,2}=\mu_{i,2}-\mu_{i,1}^2$, $\iota_{i,3}=\mu_{i,3}-3\mu_{i,2}\mu_{i,1}+2\mu_{i,1}^3$, $\iota_{i,4}=\mu_{i,4}-4\mu_{i,3}\mu_{i,1}-3\mu_{i,2}^2+12\mu_{i,2}\mu_{i,1}^2-6\mu_{i,1}^4$.

We derive the theoretical achievable rate's lower bound, when the transmit power tends to infinity as follows.
\begin{theorem}\label{Theorem_1}
When the transmit power obeys $\rho\rightarrow\infty$, the theoretical lower bound of UE-1 and UE-2, denoted as $\overline{R}_1^{(\rho=\infty)}$ and $\overline{R}_2^{(\rho=\infty)}$,  are given by
\begin{align}\label{NOMA_Design_16}
    \overline{R}_1^{(\rho\rightarrow\infty)}=\log_2\left(1+\frac{\kappa_1\varepsilon_{u^{(1)}}\varepsilon_v}
    {1-\kappa_1\varepsilon_{u^{(1)}}\varepsilon_v-\kappa_2\varepsilon_{u^{(2)}}\varepsilon_v}\right),
\end{align}
and
\begin{align}\label{NOMA_Design_17}
    \overline{R}_2^{(\rho\rightarrow\infty)}=\log_2\left(1+\frac{\kappa_2\varepsilon_{u^{(2)}}
    \varepsilon_v}{1-\kappa_2\varepsilon_{u^{(2)}}\varepsilon_v}\right),
\end{align}
respectively, which cannot increase boundlessly when $\rho\rightarrow\infty$ due to the limitation of HWI at the transceiver.
\end{theorem}
\begin{IEEEproof}
    (\ref{NOMA_Design_16}) and (\ref{NOMA_Design_17}) can be derived upon substituting $\rho\rightarrow\infty$ into (\ref{NOMA_Design_14_1}) and (\ref{NOMA_Design_14_2}).
\end{IEEEproof}

Furthermore, we derive the high SNR slope, which is a key parameter determining the scaling
law of the ergodic rate in high SNR region~\cite{zhao2022ergodic}, as follows.
\begin{remark}\label{remark_1}
According to (\ref{NOMA_Design_14_1}), the high SNR slopes of UE-1 is $S_1=\lim_{\rho\rightarrow\infty}\frac{\overline{R}_1}{\log_2\rho}=1$ when $\varepsilon_{u^{(1)}}=\varepsilon_{u^{(2)}}=\varepsilon_v=1$, and $S_1=\lim_{\rho\rightarrow\infty}\frac{\overline{R}_1}{\log_2\rho}=0$ when $\varepsilon_{u^{(1)}}<1$ or $\varepsilon_{u^{(2)}}<1$ or $\varepsilon_v<1$. By contrast, according to (\ref{NOMA_Design_14_2}), the high SNR slopes of UE-2 is $S_2=\lim_{\rho\rightarrow\infty}\frac{\overline{R}_1}{\log_2\rho}=0$ for any values of $\varepsilon_{u^{(1)}}$, $\varepsilon_{u^{(2)}}$ and $\varepsilon_v$. It means that the ergodic rate of the near-filed user, i.e. UE-1, at the high-SNR region is just limited by the HWIs at the transceivers, while that of the far-filed user, i.e. UE-2, at the high-SNR region is limited by both the HWIs at the transceivers and the date rate of the near-filed user.
\end{remark}

\begin{theorem}\label{Theorem_2}
We evaluate the theoretically achievable rate's lower bound in the case of infinitely many IOS elements as follows. When the number of IOS elements tends to infinity, the theoretical lower rate bound of UE-1 and UE-2, denoted as $\overline{R}_1^{(N\rightarrow\infty)}$ and $\overline{R}_2^{(N\rightarrow\infty)}$ respectively, tend to constant values, given by
\begin{align}\label{NOMA_Design_18}
    \notag&\overline{R}_1^{(N\rightarrow\infty)}=\log_2\Bigg(1+\\
    &\frac{\rho\varrho_1\beta_1^2 \kappa_1\varepsilon_{u^{(1)}}\varepsilon_v}
    {\rho\varrho_1\beta_1^2\left(1-\kappa_1\varepsilon_{u^{(1)}}\varepsilon_v-\kappa_2
    \varepsilon_{u^{(2)}}\varepsilon_v\right)+\eta_1^{(N\rightarrow\infty)}\sigma_{w_1}^2}\Bigg),
\end{align}
and
\begin{align}\label{NOMA_Design_19}
    \overline{R}_2^{(N\rightarrow\infty)}=\log_2\Bigg(1+\frac{\rho\varrho_2\beta_2^2
    \kappa_2\varepsilon_{u^{(2)}}\varepsilon_v}{\rho\varrho_2\beta_2^2\left(1-\kappa_2
    \varepsilon_{u^{(2)}}\varepsilon_v\right)+\eta_2^{(N\rightarrow\infty)}\sigma_{w_2}^2}\Bigg),
\end{align}
respectively, with $\eta_1^{(N\rightarrow\infty)}$ and $\eta_2^{(N\rightarrow\infty)}$ given in (\ref{NOMA_Design_20}).
\begin{figure*}[!t]
\begin{align}\label{NOMA_Design_20}
    \eta_i^{(N\rightarrow\infty)}=\frac{\left(\frac{\delta_x\delta_y}{\lambda^2}\right)^{-1}
    8\pi\left(\alpha+1\right)\left(\alpha-1\right)^2d_0^2\iota_{i,1}^2+\iota_{i,2}}
    {\frac{64\pi^2\left(\alpha+1\right)^2\left(\alpha-1\right)^4d_0^4\iota_{i,1}^4}
    {\left(\frac{\delta_x\delta_y}{\lambda^2}\right)^2}
    -\frac{16\pi\left(\alpha+1\right)\left(\alpha-1\right)^2d_0^2\iota_{i,2}\iota_{i,1}^2}
    {\frac{\delta_x\delta_y}{\lambda^2}}
    -\iota_{i,2}^2-\frac{16\left(\alpha+1\right)^2\left(\alpha-1\right)}
    {3\left(\alpha+\frac{5}{3}\right)}\iota_{i,3}\iota_{i,1}
    -\frac{\left(\frac{\delta_x\delta_y}{\lambda^2}\right)\left(\alpha+1\right)^2}
    {4\pi\left(\alpha+2\right)d_0^{2}}\iota_{i,4}}
\end{align}
\hrulefill
\end{figure*}
\end{theorem}
\begin{IEEEproof}
    When the number of IOS elements tends infinity, $A_1$, $A_2$, $A_3$ and $A_4$ can be formulated as
    \begin{align}\label{NOMA_Design_21_1}
        \notag A_1=&\left(\frac{\delta_x\delta_y}{\lambda^2}\right)^{-\frac{1}{2}}
        \int_{\infty}^{\infty}\int_{\infty}^{\infty}
        \omega^{\frac{1}{2}}\mathrm{d}x\mathrm{d}y\\
        =&\left(\frac{\delta_x\delta_y}{\lambda^2}\right)^{-\frac{1}{2}}
        \sqrt{8\pi}\left(\alpha+1\right)^{\frac{1}{2}}\left(\alpha-1\right)d_0,
    \end{align}
    \begin{align}\label{NOMA_Design_21_2}
        A_2=\int_{\infty}^{\infty}\int_{\infty}^{\infty}\omega\mathrm{d}x\mathrm{d}y=1,
    \end{align}
    \begin{align}\label{NOMA_Design_21_3}
        \notag A_3=&\left(\frac{\delta_x\delta_y}{\lambda^2}\right)^{\frac{1}{2}}
        \int_{\infty}^{\infty}\int_{\infty}^{\infty}
        \omega^{\frac{3}{2}}\mathrm{d}x\mathrm{d}y\\
        =&\left(\frac{\delta_x\delta_y}{\lambda^2}\right)^{\frac{1}{2}}
        \sqrt{\frac{2}{9\pi}}\left(\alpha+1\right)^{\frac{3}{2}}
        \left(\alpha+\frac{5}{3}\right)^{-1}d_0^{-1},
    \end{align}
    and
    \begin{align}\label{NOMA_Design_21_4}
        \notag A_4=&\left(\frac{\delta_x\delta_y}{\lambda^2}\right)
        \int_{\infty}^{\infty}\int_{\infty}^{\infty}
        \omega^{2}\mathrm{d}x\mathrm{d}y\\
        =&\left(\frac{\delta_x\delta_y}{\lambda^2}\right)
        \frac{1}{4\pi}\left(\alpha+1\right)^{2}\left(\alpha+2\right)^{-1}d_0^{-2},
    \end{align}
    respectively. Then $\eta_i^{(N\rightarrow\infty)}$ in (\ref{NOMA_Design_20}) can be derived upon substituting (\ref{NOMA_Design_21_1}), (\ref{NOMA_Design_21_2}), (\ref{NOMA_Design_21_3}) and (\ref{NOMA_Design_21_4}) into (\ref{NOMA_Design_15}).
\end{IEEEproof}

\begin{theorem}\label{Theorem_3}
For a continuous-aperture surface having $\delta_x\delta_y\ll\lambda^2$ and $N\rightarrow\infty$, the theoretical lower bounds of $\overline{R}_1$ and $\overline{R}_2$, denoted as $\overline{R}_1^{(\delta_x\delta_y\ll\lambda^2,N\rightarrow\infty)}$ and $\overline{R}_2^{(\delta_x\delta_y\ll\lambda^2,N\rightarrow\infty)}$, are given by
\begin{align}\label{NOMA_Design_22_1}
    \notag&\overline{R}_1^{(\delta_x\delta_y\ll\lambda^2,N\rightarrow\infty)}=\log_2\Bigg(1+\\
    &\frac{\rho\varrho_1\beta_1^2\kappa_1\varepsilon_{u^{(1)}}\varepsilon_v}
    {\rho\varrho_1\beta_1^2\left(1-\kappa_1\varepsilon_{u^{(1)}}\varepsilon_v
    -\kappa_2\varepsilon_{u^{(2)}}\varepsilon_v\right)
    +\frac{\left(\frac{\delta_x\delta_y}{\lambda^2}\right)\sigma_{w_1}^2}
    {8\pi\left(\alpha+1\right)\left(\alpha-1\right)^2d_0^2\mu_{1,1}^2}}\Bigg),
\end{align}
and
\begin{align}\label{NOMA_Design_22_2}
    \notag&\overline{R}_2^{(\delta_x\delta_y\ll\lambda^2,N\rightarrow\infty)}=\log_2\Bigg(1+\\
    &\frac{\rho\varrho_2\beta_2^2\kappa_2\varepsilon_{u^{(2)}}\varepsilon_v}
    {\rho\varrho_2\beta_2^2\left(1-\kappa_2\varepsilon_{u^{(2)}}\varepsilon_v\right)
    +\frac{\left(\frac{\delta_x\delta_y}{\lambda^2}\right)\sigma_{w_2}^2}
    {8\pi\left(\alpha+1\right)\left(\alpha-1\right)^2
    d_0^2\mu_{2,1}^2}}\Bigg),
\end{align}
respectively.
\end{theorem}
\begin{IEEEproof}
    According to (\ref{NOMA_Design_20}), we can show that when $\delta_x\delta_y\ll\lambda^2$, the value of $\eta_i^{(N\rightarrow\infty)}$ tends to $\frac{\frac{\delta_x\delta_y}{\lambda^2}}
    {8\pi\left(\alpha+1\right)\left(\alpha-1\right)^2d_0^2\iota_{i,1}^2}$, and (\ref{NOMA_Design_22_1}) and (\ref{NOMA_Design_22_2}) can be derived by substituting them into (\ref{NOMA_Design_18}) and (\ref{NOMA_Design_19}), respectively.
\end{IEEEproof}

\begin{figure}[!t]
    \centering
    \includegraphics[width=2.3in]{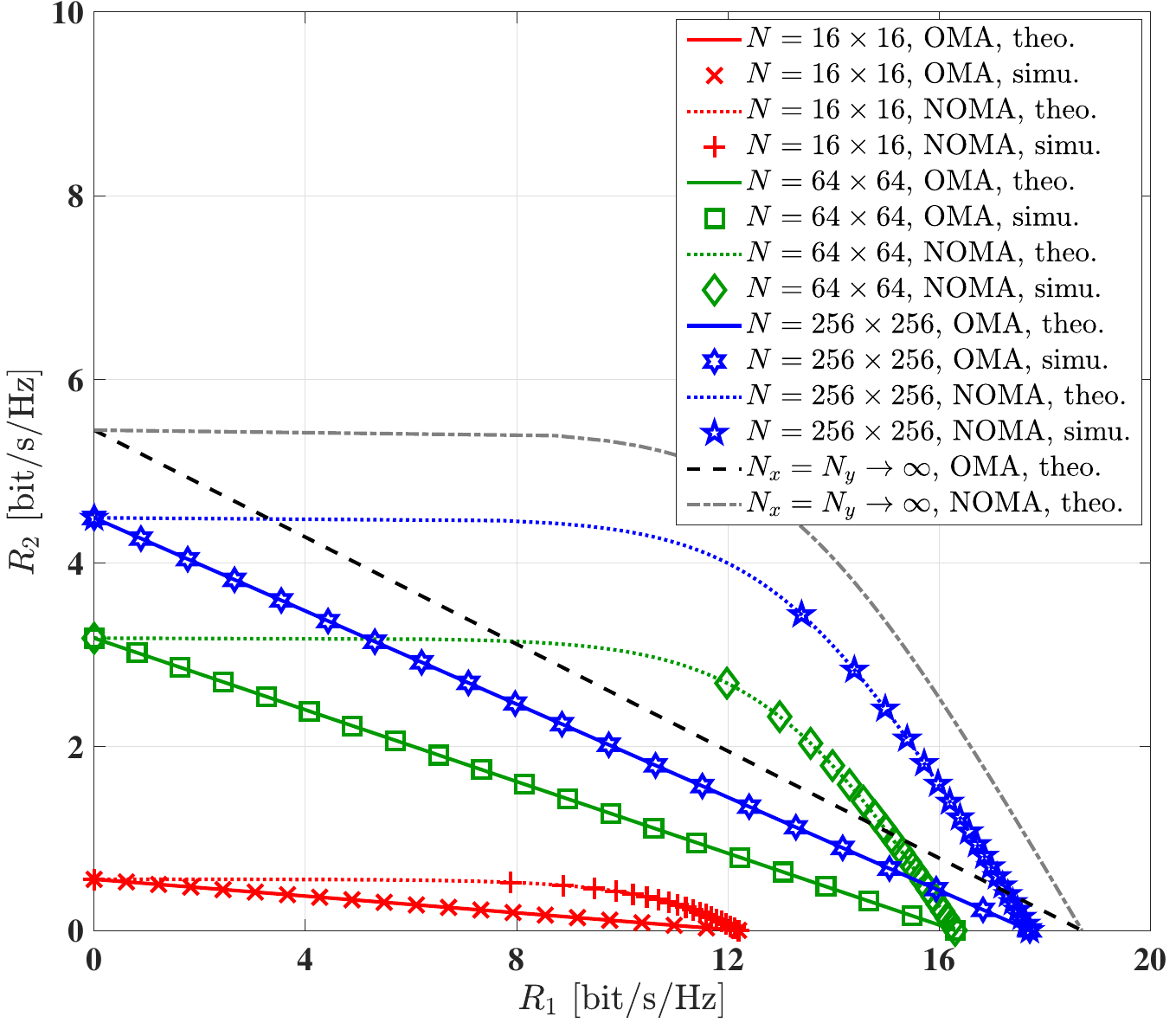}
    \caption{Theoretical lower bound and simulation results of the achievable ergodic rate comparison for UE-1 and UE-2 with different number of IOS elements.}\label{Simu_IOS_HMIMO_Fig_1}
\end{figure}

\begin{figure*}[!t]
    \centering
    \subfloat[$\varepsilon_{u^{(1)}}=\varepsilon_{u^{(2)}}=\varepsilon_v=1$]
    {\begin{minipage}{0.33\linewidth}
        \centering
        \includegraphics[width=2.3in]{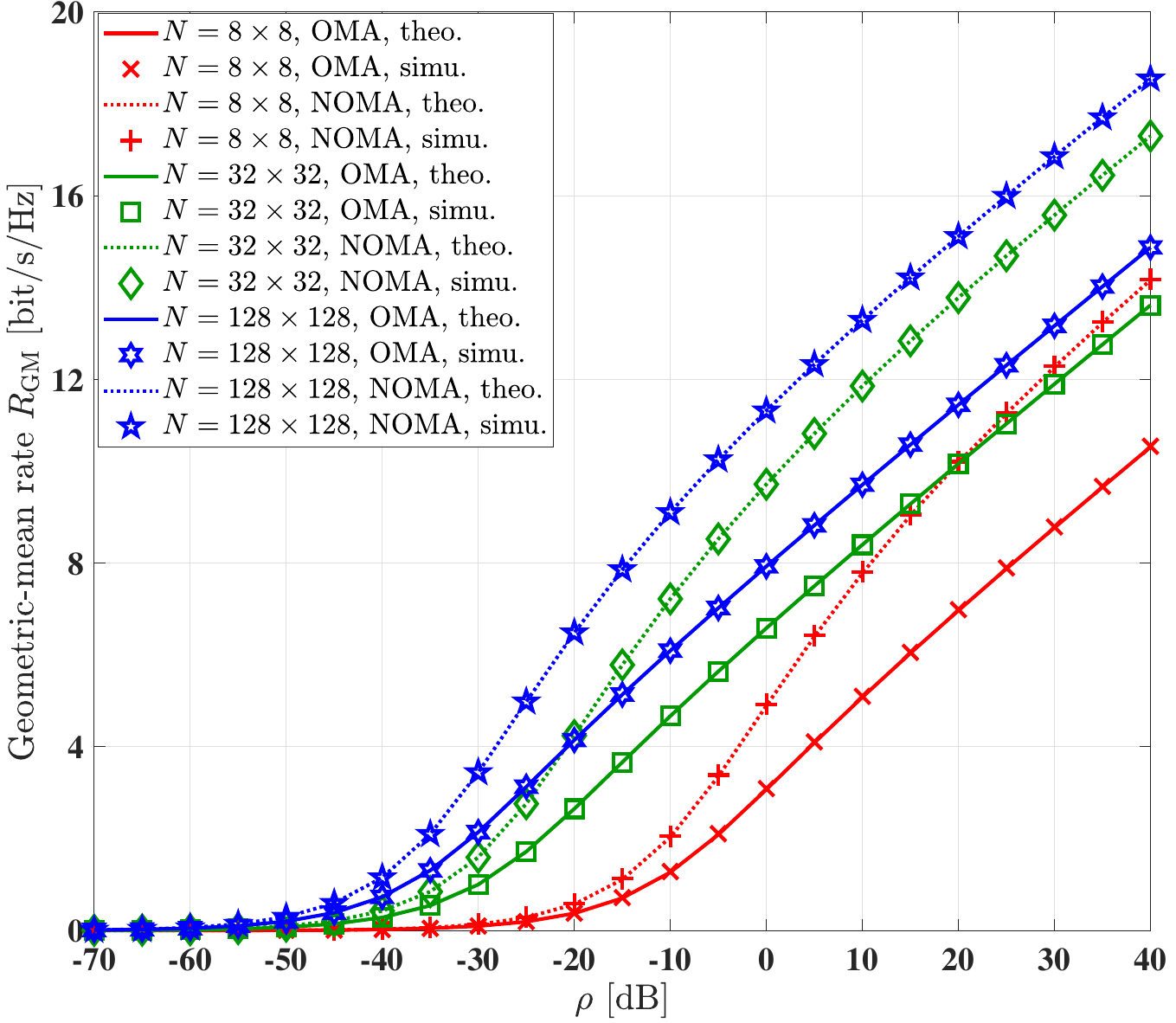}
    \end{minipage}}
    \subfloat[$\varepsilon_{u^{(1)}}=\varepsilon_{u^{(2)}}=\varepsilon_v=1-10^{-4}$]
    {\begin{minipage}{0.33\linewidth}
        \centering
        \includegraphics[width=2.3in]{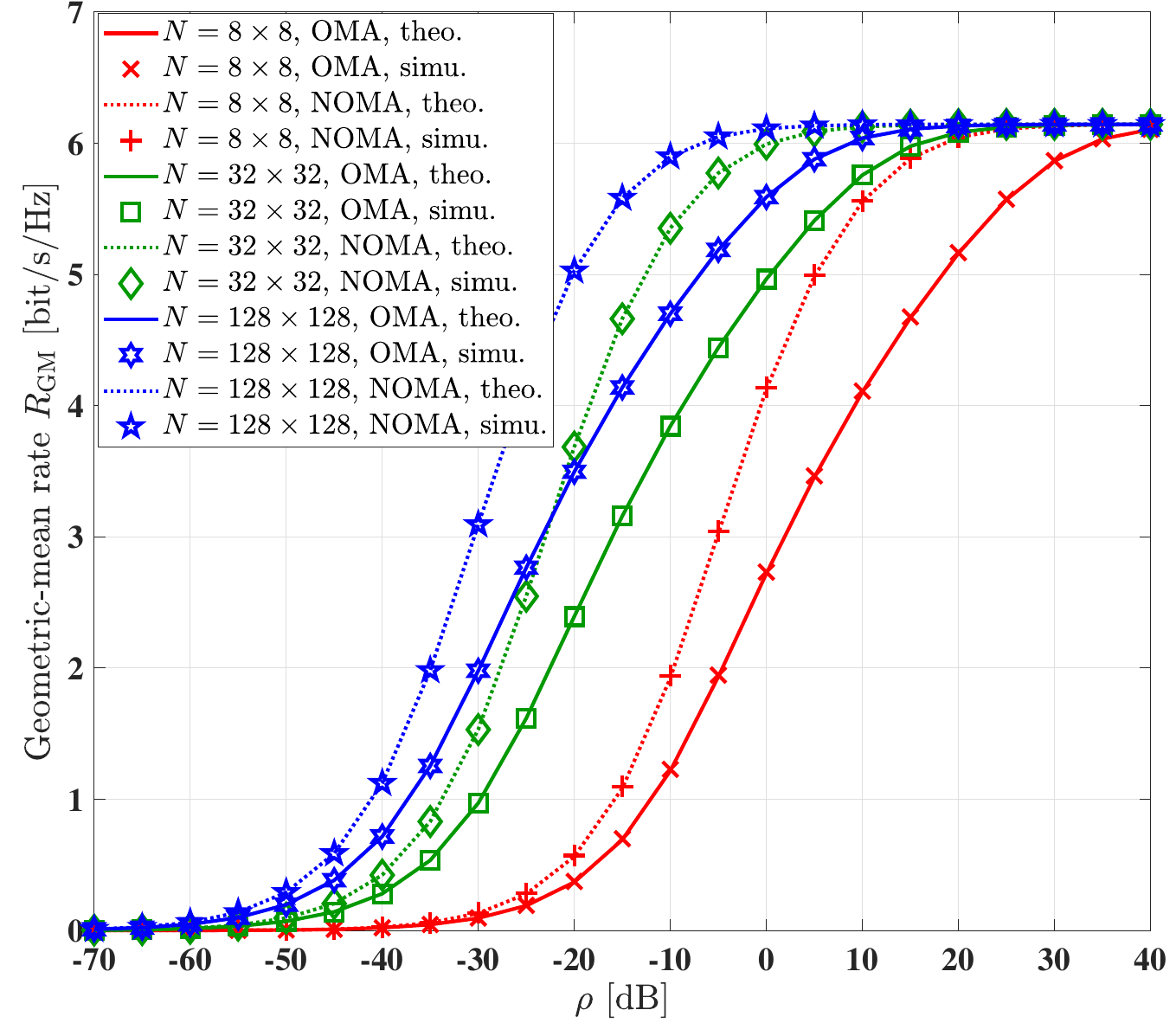}
    \end{minipage}}
    \subfloat[$\varepsilon_{u^{(1)}}=\varepsilon_{u^{(2)}}=\varepsilon_v=1-10^{-2}$]
    {\begin{minipage}{0.33\linewidth}
        \centering
        \includegraphics[width=2.3in]{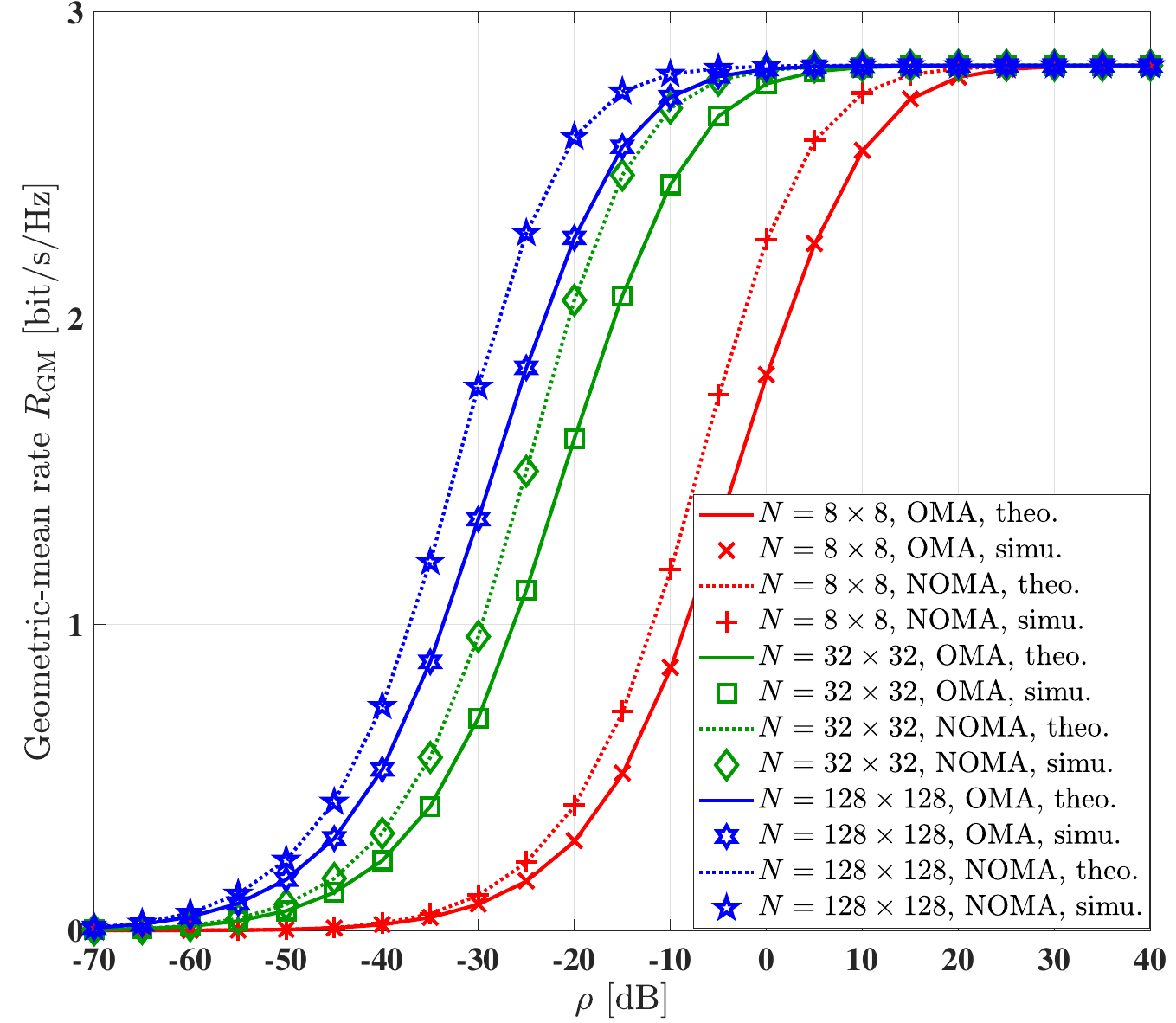}
    \end{minipage}}
    \caption{Theoretical analysis and simulation results of the geometric-mean rate $R_{\mathrm{RM}}$ with different hardware quality factors.}\label{Simu_IOS_HMIMO_Fig_2}
\end{figure*}

\section{Numerical and Simulation Results}\label{Numerical_and_Simulation_Results}
In this section, the theoretical and simulation results of the ergodic rate of the NOMA scheme are presented. They are also compared to those of the OMA scheme. The simulation parameters are: $\lambda=0.3\mathrm{m}$, $d_0=10\lambda$, $\alpha=2$, $\mu=1$. Furthermore, referring to \cite{deng2021reconfigurable_wc}, \cite{deng2021reconfigurable_tvt}, \cite{hu2022holographic}, the physical size of each IOS element equals one quarter wavelength, i.e. $\delta_x=\delta_y=\frac{\lambda}{4}$. The equal power allocation for reflecting and refracting the impinging signals on the IOS is employed~\cite{xu2021star}, i.e. $\beta_1^2=\beta_2^2=0.5$.

Both the theoretical lower bound (theo.) and the simulation results (simu.) of the ergodic rate of UE-1 and UE-2 are shown in Fig.~\ref{Simu_IOS_HMIMO_Fig_1} for different number of IOS elements in both the NOMA and OMA scheme. In this figure we have the ideal hardware quality factors of $\varepsilon_{u^{(1)}}=\varepsilon_{u^{(2)}}=\varepsilon_v=1$. The received SNRs of UE-1 and UE-2 are set to $\frac{\rho\varrho_1}{\sigma_{w,1}^2}=20\mathrm{dB}$ and $\frac{\rho\varrho_2}{\sigma_{w,2}^2}=-20\mathrm{dB}$, respectively. The achievable rate of UE-1 and UE-2 can be adjusted by controlling the parameters $\kappa_1$ and $\kappa_2$ in the NOMA scheme and $\kappa_1'$ and $\kappa_2'$ in the OMA scheme. Fig.~\ref{Simu_IOS_HMIMO_Fig_1} shows that the achievable rate is improved upon increasing the number of IOS elements $N$. Furthermore, when the number of IOS elements tends to infinity, observe in Fig.~\ref{Simu_IOS_HMIMO_Fig_1} that the achievable rates of UE-1 and UE-2 saturate at a constant value instead of increasing boundlessly due to the assumption of near-field condition for the link spanning from the feed to the IOS. This agrees with (\ref{NOMA_Design_18}) and (\ref{NOMA_Design_19}). Furthermore, the achievable rate of the NOMA scheme is higher than that of the OMA scheme.

It is meaningful to explore maximizing the geometric-mean-rate $R_{\mathrm{GM}}=\sqrt{R_1R_2}$, since it shows a substantially improved rate-fairness amongst the users \cite{yu2021maximizing}. Fig. \ref{Simu_IOS_HMIMO_Fig_2} shows the theoretical analysis and simulation results of the geometric-mean rate $R_{\mathrm{GM}}$ for different hardware quality factors. The power
allocation coefficients $\kappa_1$ and $\kappa_2$ in the NOMA scheme are optimized based on the popular bisection method, while the geometric-mean-rate in the OMA scheme is maximized when the orthogonal time/frequency resource ratio is $\kappa_1'=\kappa_2'=0.5$. The channel parameters are $\frac{\varrho_1}{\sigma_{w,1}^2}=40\mathrm{dB}$ and $\frac{\varrho_2}{\sigma_{w,2}^2}=0\mathrm{dB}$. Fig. \ref{Simu_IOS_HMIMO_Fig_2} shows that the optimized geometric-mean-rate in the NOMA scheme is better than that in the OMA scheme. When the hardware quality is ideal, i.e. $\varepsilon_{u^{(1)}}=\varepsilon_{u^{(2)}}=\varepsilon_v=1$, the geometric-mean-rate $R_{\mathrm{GM}}$ is improved without limit upon increasing the transmit power $\rho$. By contrast, when the hardware quality is non-ideal, the geometric-mean-rate $R_{\mathrm{GM}}$ becomes saturated due to the HWIs, agrees with (\ref{NOMA_Design_16}) and (\ref{NOMA_Design_17}).

\section{Conclusions}\label{Conclusion}
We conceived a HMIMO architecture based on the IOS and theoretically analyzed the ergodic rate of our NOMA scheme relying on the popular SIC detector. Both the theoretical analysis and the simulation results show that the achievable rate of the NOMA scheme is higher than that of the OMA scheme. Furthermore, the hardware impairments at the transceiver limit the achievable rate in the high-SNR region.

\bibliographystyle{IEEEtran}
\bibliography{IEEEabrv,TAMS}
\end{document}